\documentclass[sigconf,screen]{acmart}
\usepackage[utf8]{inputenc}
\usepackage{hyperref}
\usepackage{subcaption}
\usepackage{multirow}
\usepackage{booktabs}
\usepackage{cleveref}
\usepackage{enumitem}
\usepackage[skip=10pt]{caption}
\usepackage{microtype}
\usepackage[linesnumbered]{algorithm2e}

\copyrightyear{2025}
\acmYear{2025}
\setcopyright{rightsretained}
\acmConference[SC '25]{The International Conference for High Performance Computing, Networking, Storage and Analysis}{November 16--25, 2025}{St. Louis, MO, USA}
\acmBooktitle{The International Conference for High Performance Computing, Networking, Storage and Analysis (SC '25), November November 16--25, 2025, St. Louis, MO, USA}

\title{Augmenting Simulated Noisy Quantum Data Collection by Orders of Magnitude Using Pre-Trajectory Sampling with Batched Execution}

\author[T.L. Patti]{Taylor Lee Patti}
\email{tpatti@nvidia.com}
\affiliation{%
  \institution{NVIDIA}
  \country{USA}
}

\author[T. Nguyen]{Thien Nguyen}
\email{thiennguyen@nvidia.com}
\affiliation{%
  \institution{NVIDIA}
  \country{USA}
}

\author[J.G. Lietz]{Justin Gage Lietz}
\email{jlietz@nvidia.com}
\affiliation{%
  \institution{NVIDIA}
  \country{USA}
}

\author[A.J. McCaskey]{Alexander J. McCaskey}
\email{amccaskey@nvidia.com}
\affiliation{%
  \institution{NVIDIA}
  \country{USA}
}

\author[B. Khailany]{Brucek Khailany}
\email{bkhailany@nvidia.com}
\affiliation{%
  \institution{NVIDIA}
  \country{USA}
}

\definecolor{todocolor}{rgb}{0.8,0,0}
\definecolor{editcolor}{rgb}{0,0,0.8}

\newcommand{\IGNORE}[1]{}

\usepackage{listings}
\usepackage{amsfonts}
\definecolor{keywordcolor}{rgb}{0.5,0,0.5}
\definecolor{textgray}{gray}{0.4}
\definecolor{mygray}{rgb}{0.5,0.5,0.5}
\begin{abstract}
    Classically simulating quantum systems is challenging, as even noiseless $n$-qubit quantum states scale as $2^n$. The complexity of noisy quantum systems is even greater, requiring $2^n \times 2^n$-dimensional density matrices. Various approximations reduce density matrix overhead, including quantum trajectory-based methods, which instead use an ensemble of $m \ll 2^n$ noisy states. While this method is dramatically more efficient, current implementations use unoptimized sampling, redundant state preparation, and single-shot data collection. In this manuscript, we present the Pre-Trajectory Sampling technique, increasing the efficiency and utility of trajectory simulations by tailoring error types, batching sampling without redundant computation, and collecting error information. We demonstrate the effectiveness of our method with both a mature statevector simulation of a 35-qubit quantum error-correction code and a preliminary tensor network simulation of 85 qubits, yielding speedups of up to $10^6$x and $16$x, as well as generating massive datasets of one trillion and one million shots, respectively.
\end{abstract}

\begin{document}

\maketitle

\section{Introduction}

Quantum system simulation plays a pivotal role in developing quantum algorithms \cite{bharti2022nisqAlgorithms,nakaji2024generativequantumeigensolvergqe}, validating hardware performance \cite{Groszkowski_2021,chitta2022computeraidedquantizationnumericalanalysis}, and advancing quantum error correction (QEC) \cite{gidney2021stim}. As quantum processors scale, efficient classical simulations, both exact and approximate, remain essential for benchmarking hardware performance and analyzing noise effects. This is especially important due to the exponential growth in computational resources required to mathematically represent quantum states. While noiseless simulations of $n$-qubit systems demand $O(2^n)$ memory for quantum states, modeling noise via density matrices escalates this to $O(2^n \times 2^n) = O(4^n)$, rendering direct simulation intractable beyond $\sim 20$ qubits \cite{chen2020lowrankdensitymatrix,Jones2019quest}. One common method to partially overcome this computational intractability is to artificially restrict quantum circuit models to Clifford gates and Pauli noise channels, which are polynomially scaling groups \cite{gottesman1998heisenberg}. This technique, which is used in software packages such as Stim \cite{gidney2021stim}, enables the simulation of large quantum systems, but is not capable of simulating general (universal) quantum computers.

To address this gap between classical tractability and quantum accuracy, quantum trajectory methods can be used to approximate density matrices through statistical ensembles of $m \ll 2^n$ noisy statevectors \cite{dum1992,dalibard1992,daley2014}. This approach achieves quadratic memory savings on exponentially-scaling data structures ($2^n$ vs $2^n \times 2^n$) by propagating stochastic noise injections rather than explicit matrix representations. CUDA-Q’s GPU-accelerated simulator facilitates quantum trajectory methods, and further optimizes them through batched trajectory execution, unitary-channel detection for probability caching, and distributed multi-GPU architectures \cite{bayraktar2023cuquantumsdkhighperformancelibrary}. Despite these advances, current trajectory methods face three fundamental limitations: (1) unoptimized sampling requiring redundant state preparations per trajectory, (2) single-shot statistical data collection obscuring error provenance, and (3) rigid sampling strategies ill-suited for targeted error analysis in QEC and machine learning applications.

\begin{figure*}
\includegraphics[width=\textwidth]{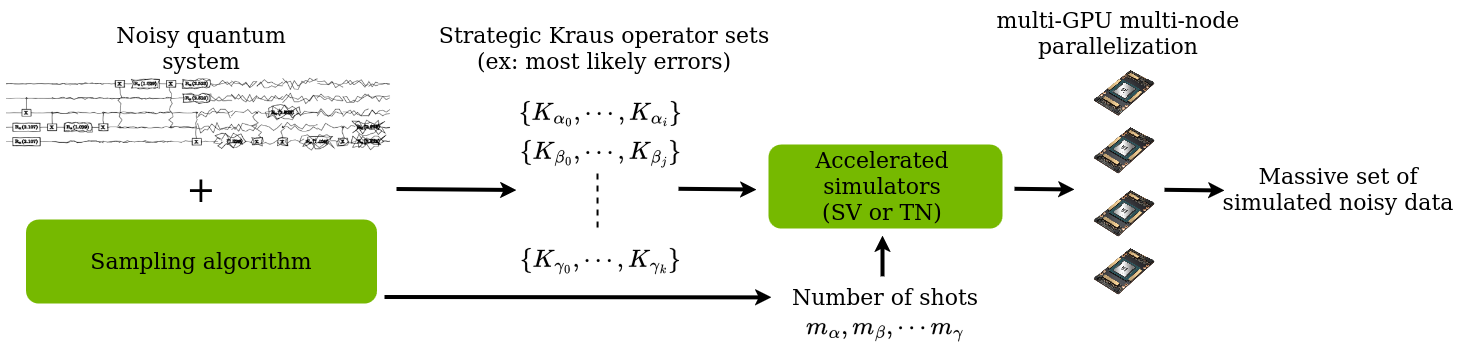}
    \hfill
    \caption{\textbf{Diagram of PTSBE:} An arbitrary noisy circuit is registered into the desired sampling algorithm for PTS, such as one that samples the Kraus operator space proportionally, exhausts the most likely error combinations, etc. The prescribed sampled sets of Kraus operators $ \{ K_{\alpha_0}, \cdots, K_{\alpha_i} \} $, along with their prescribed number of shots $m_\alpha$, are passed from the PTS algorithm to the CUDA-Q simulator using either a statevector or tensor network backend. This enables the Kraus operator sets to be strategically sampled through a quantum measurement-like process, with the desired number of measurement shots collected in bulk without quantum state repreparation. Both CUDA-Q backends can be distributed across multiple GPUs, both within a single trajectory and for simultaneous trajectories in an embarrassingly parallel manner, permitting faster and larger simulations.}
    \label{fig:ptsa}
\end{figure*}

In this work, we advance quantum trajectory methods by introducing Pre-Trajectory Sampling (PTS) with Batched Execution (BE), or PTSBE, a methodology that decouples stochastic noise sampling from quantum state evolution to overcome the aforementioned bottlenecks. Conventional trajectory simulations interleave gate applications with per-step noise sampling, forcing serial execution and discarding critical error metadata. In contrast, PTS precomputes all stochastic decisions – including error types, locations, and mitigation parameters – prior to statevector propagation. This enables three key innovations:

\begin{itemize}
    \item Tailored error injection for specific QEC analysis scenarios (e.g., Pauli twirling or spatially correlated noise)
    \item Batched trajectory execution (BE) eliminating redundant circuit recompilation and state initialization
    \item Error provenance tracking through lightweight metadata tags attached to each trajectory
\end{itemize}

We demonstrate PTSBE’s effectiveness via two implementations: a mature 35-qubit QEC code simulation achieving $10^6\times$ speedup over conventional trajectory methods on GPU clusters, and a preliminary 85-qubit QEC code preparation circuit tensor network simulation showing $16\times$ acceleration. This considerable acceleration enabled us to produce noisy quantum datasets, such as would be useful for training AI-based QEC decoders \cite{Chamberland2018deepneuraldecoders,Baireuther_2019,Sweke_2020}, a crucial application of AI for quantum science \cite{alexeev2024artificial}. The datasets generated included a one trillion-shot, 35-qubit statevector simulation and one-million shot, 85-qubit tensor network simulation. Using PTSBE methods, these extremely large datasets can be obtained with unprecedented speed, requiring just $4{,}445$ H100 GPU hours and $2{,}223$ H100 GPU hours on NVIDIA's Eos DGX Superpod \cite{eos} when simulating $10^6$ statevector shots and $100$ tensor network shots per trajectory, respectively. These results highlight PTSBE’s potential to enable practical simulation of scalable QEC architectures and generate training data for ML-based error decoders, such as have been targeted in Google's recent AlphaQubit work \cite{google2024alphaqubit}, which used moderately-sized experimental datasets and massive Clifford simulator-based datasets. PTSBE vies to simplify and expand this training paradigm by supplementing more traditional data modalities with large amounts of data from universal quantum simulators. By transforming trajectory simulation from a statistical black box into a programmable data collection engine, this work bridges critical gaps in noisy quantum system analysis.

\section{Background}\label{sec:background}

\subsection{Simulating Noisy Quantum Systems}\label{sec:noisy-quantum}

Quantum states that have not been subjected to noise or are otherwise part of a probabilistic ensemble are referred to as ``pure'' \cite{shankar2012principles}. Such pure states can be described by a $2^n$-element one-dimensional complex vector, where $n$ is the number of two-level quantum systems or ``qubits'' as they are often called \cite{nielsen2010quantum,gill2022quantum}. Quantum formalism dictates that when a state is measured, e.g., in the laboratory, its $2^n$ entries are collapsed into stochastic ``shot'' values of length $n$ via a probabilistic sampling-like process \cite{jordan2024quantum}. As these shot values, or quantum measurements, are the experimentally observable data for quantum systems, they are commonly the data reproduced by quantum simulators, with the $2^n$-dimensional state vectors first being prepared and then sampled for some number of $n$-dimensional shot values. Once the exponentially complex quantum state is prepared, repeated shot sampling is typically very efficient, scaling in just polynomial time for well-optimized statevector simulations \cite{cuda-q}.

While pure states are useful for studying a wide array of fundamental and idealized phenomena, modeling experimentally realistic and naturally occurring quantum systems requires a more complex representation \cite{scully1997quantum,campaioli2024quantum}. The most complete of these representations is the systems density matrix, a $2^n \times 2^n$-dimensional complex matrix that provides the full description of a noisy or mixed quantum system \cite{shankar2012principles}. This added factor of $2^n$ adds substantially to the complexity of simulating the quantum system at hand, greatly reducing the number of qubits that can be simulated without approximation.

Common simplifications of the full density matrix formalism include approximate tensor networks (where the richness of the quantum distribution is reduced by tensor truncation) \cite{white1992density,baiardi2019large}, extended Clifford gate simulator approximations for open quantum systems (where the majority of gates in a quantum system are approximated to come from a tractably small, closed group) \cite{gottesman1998heisenberg,bravyi2016improved}, approximate reduction of the system itself \cite{schrieffer1966relation,brion2007adiabatic,yadav2023legate,chakraborty2024gpu}, and quantum trajectory techniques. Quantum trajectory techniques are the method used in this paper and the subject of the following subsection.

\begin{figure}
\includegraphics[width=\linewidth]{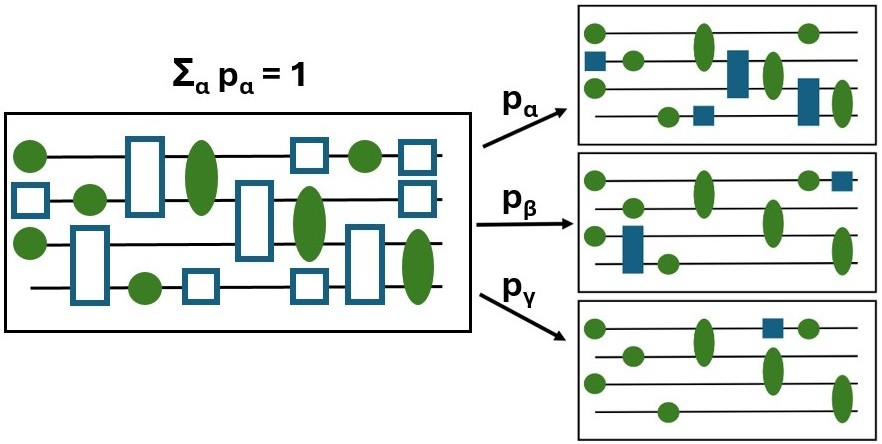}
    \hfill
    \caption{The partitioning of a noisy quantum circuit into sampled sets of Kraus operators. The full noisy circuit (left) contains both coherent gates (solid green circles/ovals) and noisy operations (empty blue squares/rectangles). A sampling algorithm selects subsets of Kraus operators (right, solid blue squares/rectangles, $ \{ K_{\alpha_0}, K_{\alpha_1},
    K_{\alpha_2},
    K_{\alpha_3} \} $, $ \{ K_{\beta_0}, K_{\beta_1} \} $, and $ \{ K_{\gamma_0} \} $) from which to sample shots. The full distribution of Kraus operator subsets $\{ p_\alpha \}$ has unit probability, although typically only a subset of these will be sampled.}
    \label{fig:partitioning}
\end{figure}

\subsection{Traditional Trajectory Simulations with CUDA-Q}\label{sec:trajsim-CUDA-Q}

In CUDA-Q~\cite{cuda-q}, the GPU-accelerated state vector and tensor network simulator backends, indexed as \texttt{nvidia} and \texttt{tensornet}, respectively, can simulate quantum noise via the trajectory method. Specifically, given a quantum noise channel represented by a set of Kraus operators $\{K_i\}$ satisfying the completely positive and trace preserving (CPTP) condition~\cite{nielsen2010quantum}, trajectory methods can stochastically simulate the effect of this noise channel on the quantum system by randomly selecting samples of $K_i$ to apply to ensembles of quantum states with probability $\langle\psi|K_i^\dagger K_i|\psi\rangle$~\cite{breuer2002theory, wiseman2009quantum}. Due to the ensemble nature of this technique, a considerable number $m$ of trajectories must be simulated for an accurate distribution of quantum shots. However, for many systems $m \ll 2^n$, rendering trajectory methods a strong alternative to traditional density matrix simulations. It is worth noting that the CPTP condition guarantees that the sum of probabilities is always unity for arbitrary quantum states. 

Compared to the density matrix simulation approach, the quantum trajectory method allows us to use the state vector representation, which is quadratically smaller ($2^n$ for state vectors vs. $2^n \times 2^n$ for density matrices); enabling us to simulate twice as many qubits. The disadvantage to this method is that we must simulate and sample shots from a large number of distinct noisy trajectories independently in order to collect a representative statistical ensemble, e.g., an accurate noisy shot distribution.

We implement PTSBE using the CUDA-Q trajectory noisy simulator.  Before describing PTSBE enhancements in Sec.\ \ref{sec:ptsa}, we first introduce key features in the baseline CUDA-Q framework used for simulating quantum trajectory methods. The traditional CUDA-Q trajectory noisy simulator has the following set of pre-existing features that are novel among quantum simulators and are not present in previous works, such as~\cite{dd_densitymat, intel_qs, qsim, dmsim, mpdo_sim, gidney2021stim, dd_noisy_sim, aer_trajectory}:

\begin{enumerate}
\item \textbf{Support for single-GPU and multi-node multi-GPU simulations:} Users can scale the maximum number of simulatable qubits, or the speed of their simulations, by deploying additional GPUs. This multi-GPU capability is also harnessed by our PTSBE method.

\item \textbf{Efficient simulation of unitary mixture noise channels:} Noise channels, as specified by their set of Kraus matrices, are automatically analyzed to detect whether they are unitary-mixture channels, i.e., $K_i = p_i U_i, \forall K_i$, where $U_i$ is a unitary matrix. For example, the commonly used depolarizing channel is a unitary mixture of Pauli unitaries \cite{nielsen2010quantum}. Unitary channels can be simulated more efficiently as the probability is quantum-state independent. We emphasize  that, outside of the innovation of this work, CUDA-Q executes all trajectories independently. That is, the distinction between unitary-mixture and general (Kraus-based) channels only affects how each trajectory is simulated, and this distinction is fully different from the efficient trajectory sampling mechanism introduced in this manuscript.    

\item \textbf{Batching of trajectory simulations for small systems:} This feature is only available on the state vector simulator. When the number of qubits is small, the backend can batch multiple, independent trajectory simulations in the same run in an embarrassingly parallel manner in order to maximize the GPU utilization and reduce the simulation time.
We again emphasize that this GPU-level batching mechanism is different from the PTSBE workflow proposed in this manuscript as the latter: 1) focuses on the simulation of large systems for which the traditional batching would be unsuitable, 2) batches quantum shot evaluation to eliminate duplicate work and generate massive amounts of data, 3) provides users control over the types of noisy data collected, and 4) provides users with data on error providence.

\end{enumerate}

The workflow for traditional noisy trajectory simulation using the current CUDA-Q backend is shown in Algorithm~\ref{alg:traj_sim}, We there illustrate the high-level simulation workflow differences between unitary mixture and general noise channels. In particular, the latter requires expectation calculation based on the current state at the trajectory sampling point. The CUDA-Q trajectory simulator uses the CUDA random number generation library (cuRAND) to generate random numbers for optimal performance.

Transparent to this algorithm is the distributed state vector capability, supporting multi-node multi-GPU, of the simulator. In the large-scale simulation mode, the gate matrix application and expectation value calculation steps are also distributed, operating on slices of the state vectors and consolidating the results.

\begin{algorithm}
\caption{\textbf{Pseudocode block for traditional noisy trajectory simulation on the current CUDA-Q backend. While there is some trajectory-based optimization in the form of discerning between unitary and Kraus-based error formalisms, the advantages furnished by PTSBE have not been achieved prior to our work.}}\label{alg:traj_sim}
\KwData{\texttt{noiseModel}, \texttt{operatorSequence}}

\For{\texttt{operator} \texttt{in} \texttt{operatorSequence}}{
  apply \texttt{gate} matrix\;
  $\texttt{noiseChannel} \gets \texttt{lookUp}(\texttt{noiseModel}, \texttt{operator})$\;
  $r  \gets \texttt{cuRAND()}$ \;
 \eIf{\texttt{noiseChannel} \texttt{is} \texttt{unitaryMixture}}{
    $k = \texttt{index}(r, \{p_i\})$ \;
    $\texttt{applyMatrix}(U_k)$ \;
  }{
    $p_i  \gets \langle \Psi | K_i^\dagger K | \psi \rangle$\;
    $k = \texttt{index}(r, \{p_i\})$ \;
    $\texttt{applyMatrix}(K_k /\sqrt{p_k})$ \;
  } 
}
\end{algorithm}

\subsection{PTSBE Application: Massive Data Collection for Quantum Error Correction}
\label{sec:qec}

One widespread application of noisy quantum simulators is to characterize the deliterious effects of noise on gate-based quantum computers \cite{georgopoulos2021NISQmodeling}. The sources of these nonidealities are varied, ranging from engineering concerns (e.g., imperfect quantum gate application \cite{bartolomeo2023noisygates,bharti2022nisqAlgorithms}), to the fundamental challenges of open quantum systems (e.g., undesirable environmental coupling \cite{breuer2002theory,nielsen2010quantum}). In particular, noisy simulators play an important role in the study of quantum error correcting codes. QEC is a set of tools used to combat the effect of noise on the fragile information that quantum computers store and process, enabling quantum computers to perform calculations fault-tolerantly \cite{steane2003threshold,fowler2009high_threshold}.

Much like their classical error-correction counterparts, many approaches to QEC encode the information used and generated during a quantum algorithm in a smaller logical subspace of the larger (Hilbert) space formed by the qubits \cite{google2023logicalqubit,google2024qecbelowthreshold,ryananderson2022implementingfaulttolerantentanglinggates}. Once the information is encoded, stabilizer measurements can be performed to query parity-check information about the quantum state. These parity-check measurements do not collapse the logical quantum state, but give indirect information about what error mechanisms have corrupted the logical information \cite{nielsen2010quantum}. This process of using stabilizer measurements to infer what errors occur is called decoding and can itself be a computationally expensive task \cite{iyer2013hardnessdecodingquantumstabilizer}, \cite{Battistel2023decodingchallenges}. 

Noisy simulation of QEC circuits provides considerable research value, as knowledge of underlying error origin and propagation can validate a variety of QEC protocols. Once the decoder attempts to infer which error mechanisms have occurred, a simulator can directly validate this output and provide the decoder with either correction or reinforcement.

As QEC circuits encode logical qubits into a greater number of physical qubits, QEC simulations have an inherently larger classical memory requirement. This additional overhead is particularly vexing given that each additional qubit added to the circuit quadruples the space requirements for density matrix simulation of the noisy system. These limitations have led to Clifford simulation emerging as the most common simulation tool in the QEC research space, as this simulation strategy has very modest time and memory complexity \cite{gottesman1998heisenberg}. Clifford simulation does have a distinct disadvantage in that it does not admit a universal gate set, limiting its applicability and modeling power. That being said, there are various protocols within QEC that are comprised entirely of Clifford gates, and Clifford gate-based software packages such as Stim \cite{gidney2021stim} have proven extraordinarily useful for the study of a variety of QEC codes in noisy circuit contexts. More specifically, by restricting itself to Clifford circuits and Pauli noise channels, Stim is able to use a reference frame sampler to efficiently bulk sample noisy simulation data at a rate of MHz. However, the poignant limitations of Clifford simulations remain, necessitating the development of highly-efficient methods of noisy, universal quantum data collection, such as PTSBE.

\begin{figure}
    \centering
    \includegraphics[width=0.9\linewidth]{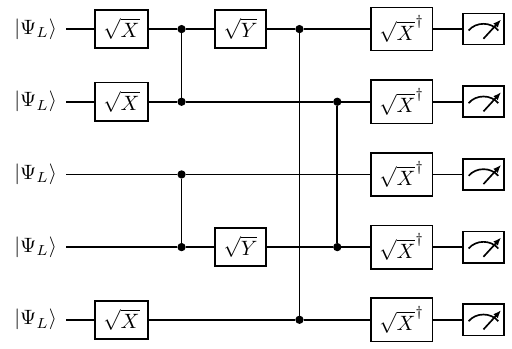}
    \caption{Quantum circuit diagram depicting the 5-1 magic state distillation protocol as compiled described by\cite{quera2024logicalMSD}. In this case, each qubit wire represents either 7 or 17 physical qubits encoding a single logical qubit, which corresponds to the 35 and 85-qubit simulations demonstrated in this work. Here the top wire was measured in all three Pauli bases so that the fidelity of the resulting magic state could be computed.}
    \label{fig:msd_diagram}
\end{figure}

In this work, we use PTSBE to study the 5-1 magic state distillation (MSD) protocol \cite{Bravyi2005MSD} that was recently implemented on QuEra’s neutral atom quantum computer \cite{quera2024logicalMSD} and depicted in Fig.\ \ref{fig:msd_diagram}. Although this was not the first experimental realization of MSD, it was the first to encode such a data set in logical qubits, here in the [[7,1,3]] color code, and the [[17,1,5]] color code. While 5 qubits are easily simulated by a variety of frameworks, this protocol uses 5 \textit{logical qubits} resulting in a total of 35 (7-qubit code) or 85 (17-qubit code) \textit{physical qubits}. While 35 and 85 qubits are also easily simulated by Clifford simulators, MSD is a QEC protocol that consumes and produces non-Clifford states, making universal noisy quantum simulation methods like PTSBE highly beneficial.

Indeed, PTSBE, while applicable to nearly any application of large-scale quantum trajectory simulation, was developed in order to generate synthetic training data for an AI decoder. While AI decoders show promise to overcome accuracy and computational challenges of decoding, these methods require high-quality training data \cite{Baireuther_2019,Sweke_2020,Chamberland2018deepneuraldecoders} and, while data supplied by a physical quantum computer would be most desirable, it's extremely expensive at scale. In Google's recent AlphaQubit work \cite{google2024alphaqubit}, moderately-sized experimental datasets and massive Clifford simulator-based datasets were used to train an AI decoder, a training paradigm that we hope to simplify and expand by supplementing more traditional data modalities with large amounts of data from universal quantum simulators. For the purpose of training AI decoders, this training data has an additional benefit of known error providence. This information can be used as training labels on the output data to enable supervised learning, which is not possible for data derived from quantum devices and is not a feature that was previously available for trajectory simulators.

\section{Pre-Trajectory Sampling with Batched Execution for Noisy Quantum Systems}\label{sec:ptsa}

PTSBE methods separate noisy quantum trajectory sampling out into two steps: 1) the strategic sampling of desired sets of Kraus operators (PTS) and 2) the preparation and batched shot execution on the resulting quantum states (BE). The former of these two steps allows us to be efficient and judicious in the potential errors from which we sample, whereas the latter allows us to obtain massive datasets from the sampled error sets without redundant state preparation.

Further detail on PTSBE methods is given in Fig. \ref{fig:ptsa}. The user specifies a noisy quantum system of interest by inputting the quantum operations, both coherent and thus deterministic (solid green circles/ovals in Fig. \ref{fig:partitioning}) and noisy and thus randomly sampled (hollow blue squares/rectangles in Fig. \ref{fig:partitioning}). A sampling algorithm of the user's choice is then used to select subsets of noisy operations $\{K_{\alpha_0} \cdots K_{\alpha_i} \}$ (solid blue squares/rectangles in Fig. \ref{fig:partitioning}) from the total distribution of the noisy quantum system, along with the appropriate number of shots $m_\alpha$ to be sampled for each Kraus operator. Unlike traditional black box quantum trajectory methods, PTS saves crucial yet lightweight metadata, such as the Kraus operator errors $\{K_{\alpha_0} \cdots K_{\alpha_i} \}$, which can be useful in downstream tasks (e.g., as training data labels). The details of a given PTS algorithm are chosen to provide Kraus operator sets that are strategic for the usecase at hand, as detailed in Sec. \ref{sec:sampling_algorithms}.

After PTS, each set of Kraus operators $\alpha$ is then loaded into a CUDA-Q simulator for BE. CUDA-Q prepares the state obtained from both the coherent gates and the now fixed noise gates \\
$\{K_{\alpha_0} \cdots K_{\alpha_i} \}$. As the state is of $O(2^n)$ complexity to prepare and store, a major benefit of BE is that it avoids redundant state preparation by sampling all $m_\alpha$ desired quantum bitstrings (measurements) at once, a task of mere polynomial complexity. PTS methods are agnostic to simulator design, and we demonstrate its performance using both statevector and tensor network BE implementations, as detailed in Sec. \ref{sec:benchmarking} and Figs.\ \ref{fig:sv_benchmarking} and \ref{fig:tn_benchmarking}. In order to fit larger quantum systems in memory, or to accelerate their preparation and sampling, PTSBE can be parallelized over multiple GPUs. This is true for both intra-trajectory preparation (e.g., deploying more GPUs on the preparation of a single quantum state), as well as inter-trajectory preparation (as the preparation and sampling of different trajectories is embarrassingly parallel, the calculation process trivially scales to arbitrarily many GPUs). This batched efficiency and multi-faceted parallelization permits the collection of massive corpuses of quantum data, such as those used to characterize large noisy quantum systems or to train AI models for quantum dynamics (as discussed in Sec.\ \ref{sec:qec}).

\subsection{Pre-Trajectory Sampling Algorithms}\label{sec:sampling_algorithms}

The basis of PTSBE methods are the PTS algorithms that pre-select the Kraus operators and number of quantum shots to be obtained by noisy quantum trajectory simulation with batched execution. This separates the lightweight Kraus operator selection process from the computationally-intensive state preparation and state sampling processes, enabling the latter two to be done in a batched and/or parallel manner. Likewise, PTS algorithms inform the user of the providence of errors by providing their details as lightweight metadata, as well as enable the user to target a subset of Kraus operator error combinations, such as error combinations with certain ranges of likelihood or those with certain characteristics. This further distinguishes PTSBE from traditional noisy quantum trajectory sampling, which are less transparent and more restrictive in their sampling approach.

Algorithm \ref{alg:probabilistic_sampling} is a simple example of a PTS algorithm. It probabilistically samples a specified number of times ($\texttt{nsamples}$) from the distribution $\texttt{NoisyCircuit}(\{K\}, \{p\})$ of Kraus operators $\{K\}$ with probabilities $\{p\}$, requiring just $\sim O(|\{K\}|^2 (\overline{p})^2)$ operations. Physically incompatible Kraus error combinations, such as two operators that would act on the same qubit at the same time are removed ($\texttt{compatible}$ function), as are duplicate $\texttt{KrausSample}$ trajectories ($\texttt{uniqueKraus}$ function). In this example, a large, uniform number of shots ($\texttt{nshots}$) is assigned to each set of Kraus operator errors in order to maximize data collection, such as would be useful for training ML models for quantum applications (e.g., QEC decoders as discussed in Sec. \ref{sec:qec}).

Numerous straightforward expansions on Algorithm \ref{alg:probabilistic_sampling} can be constructed to facilitate a wide variety of sampling tasks. For example, if the user desires a more proportionally sampled dataset, e.g., for expectation value estimation, they can achieve this by using the error probabilities $p$ for each $K$ to calculate joint probability $p_\alpha$ of each $\texttt{KrausSample}$ $\{K_{\alpha_0} \cdots K_{\alpha_i} \}$ and then redistributing or resampling the number of shots allocated to each Kraus operator set $\{K_{\alpha_0} \cdots K_{\alpha_i} \}$ according to the relative populations $\{p_\alpha'\}$, where $p_\alpha' = p_\alpha / \sum_i p_i$. Such variations also support preferred sampling from probability bands, wherein a Kraus operator set $\{K_{\alpha_0} \cdots K_{\alpha_i} \}$ is only chosen if $p_\alpha \in [p_\text{min}, p_\text{max}]$. Similarly, the most common errors can be calculated analytically by considering only error combinations whose joint probability falls above a given cutoff, a combinatorial problem of generally tractable order when considering experimentally relevant noise probabilities and sizeable error cutoffs. Separately, we could also add selection criteria to $\texttt{Line 5}$ of Algorithm $\ref{alg:probabilistic_sampling}$ to specify gate type, parity, location, and so on. 

\begin{algorithm}
\caption{\textbf{Pseudocode for a basic probabilistic PTS algorithm. The Kraus operator sets $\{K\}$ with probabilities $\{p\}$ are sampled from the full distribution of $\texttt{NoisyCircuit}$ in much the same way as they would be at state preparation runtime in a traditional simulator, but avoiding redundant, computationally expensive state preparation by avoiding sampling of repeated Kraus operator sets $\texttt{KrausSample}$ and pre-assigning a large number of shots $\texttt{nshots}$ for collection from each state.}}
\label{alg:probabilistic_sampling}
$\textbf{Inputs: } \texttt{NoisyCircuit}(\{K\}, \{p\}) \texttt{, nsamples, }\texttt{nshots}$ \\
$\texttt{KrausSets, KrausShots} = []\texttt{, } []$ \\
\For{\texttt{sample in range}(\texttt{nsamples})}{
$\texttt{KrausSample} = []$ \;
  \For{$K, p \texttt{ in NoisyCircuit}(\{K\}, \{p\})$}{
  $r \gets \texttt{randomUniform}(0, 1)$\;
    \If{$r \leq p$}{
       \If{$\texttt{compatible}(K, \texttt{KrausSample})$}{
         $\texttt{KrausSample.append}(K)$;
       }
    }
  }
  \If{$\texttt{uniqueKraus}(\texttt{KrausSample, KrausSets})$}{
    $\texttt{KrausSets.append(KrausSample)}$ \;
    $\texttt{KrausShots.append(nshots)}$ \;
  }
}
$\textbf{Returns: } \texttt{KrausSets, KrausShots}$
\end{algorithm}

\section{Scaling Experiments}\label{sec:benchmarking}

\begin{figure}
\includegraphics[width=\linewidth]{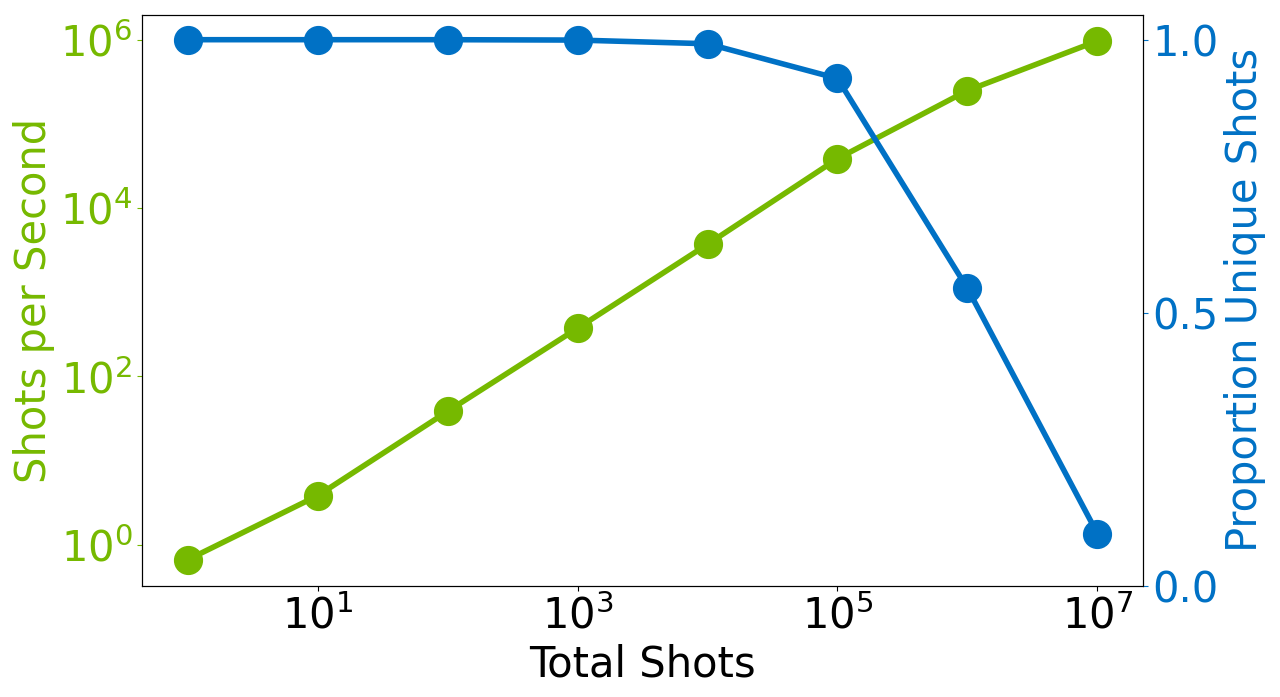}
\caption{The number of shots collected per second (green, left axis) and the proportion of unique shots collected (blue, right axis) as a function of the total shots sampled for a 35-qubit magic state distillation circuit using statevector simulation. As shot sampling is much more efficient than statevector preparation, sampling large numbers of shots from a given Kraus operator subset is highly efficient and allows for the collection of a massive data corpus. Due to the intricacies of the $2^{35}$-dimensional statevectors prepared, the additional shot data collected remains extremely useful even for very large sample sizes, with samples upwards of $10^6$ shots being comprised of largely unique information.}
\label{fig:sv_benchmarking}
\end{figure}

PTSBE methods demonstrate marked speedup for both statevector and tensor network backends. This is especially true for the statevector implementation (Fig.\ \ref{fig:sv_benchmarking}), as the tools required for carrying out efficient batched sampling were largely available in the CUDA-Q library. To benchmark this implementation, we used the 35-qubit magic state distillation circuit discussed in Sec.\ \ref{sec:qec}. Four H100 GPUs with 80GBs of vRAM each were used in each trajectory simulation, as this was the minimum number able to accomodate the sizeable memory footprint of a 35-qubit statevector, which contains $2^{n+1} \texttt{ float32}$ values (i.e., $2^n \texttt{ complex64}$ values). As sampling quantum shots from a statevector takes a fraction of the time of statevector preparation itself, the increase in shots per second for batched sampling is nearly linear with the total number of shots collected per Kraus operator set (left axis, green), with the efficiency increase reaching $\sim 10^6$ for batch sizes of $10^6{-}10^7$ shots. As the information stored within such large quantum states is so sizable, the shots sampled remain largely distinct and thus generally contribute novel information, even in the regime of extremely large batch sizes. Indeed, samples of $10^6$ total shots are comprised of more than a $0.5$ fraction of unique results (right axis, blue).

\begin{figure}
\includegraphics[width=\linewidth]{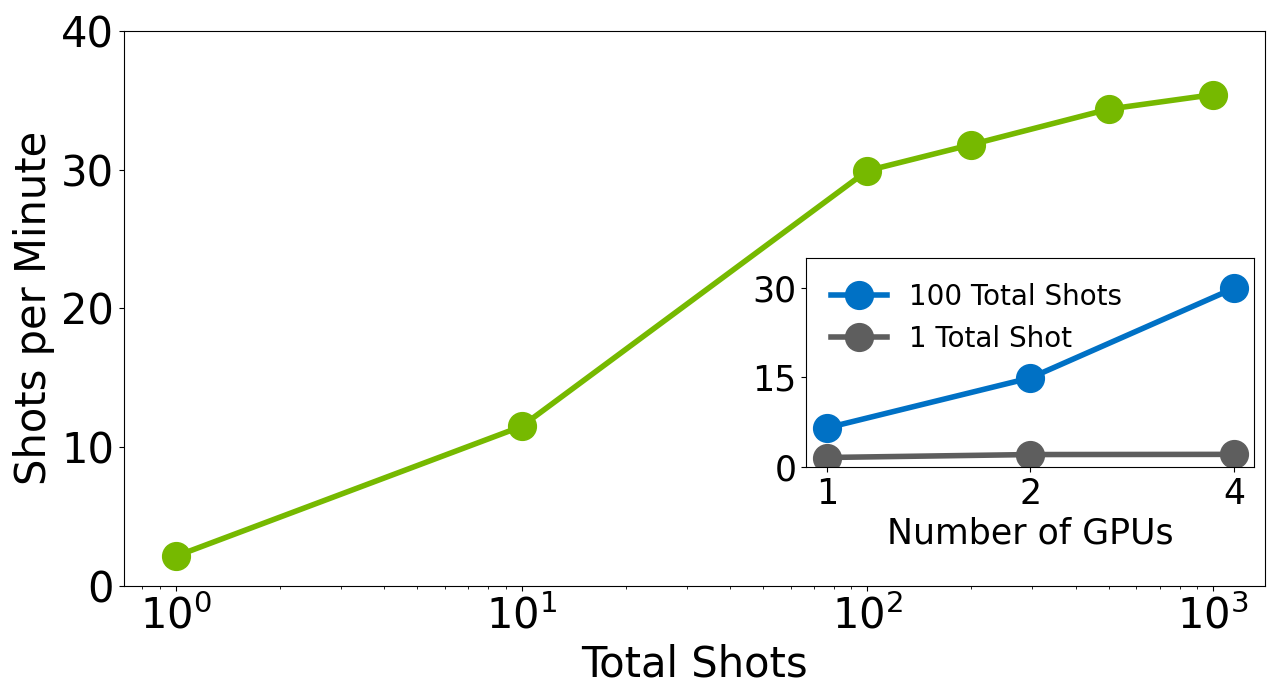}
\caption{The number of shots collected per minute as a function of the total shots sampled for an 85-qubit magic state distillation circuit using tensor network simulation. Repeated shot sampling can be considerably more efficient than full tensor network preparation, leading to a substantial increase in efficiency with increased shot number. That being said, the efficiency of multi-shot executions are highly dependent on tensor network contraction path caching and adaptation, as well as efficient correlated sampling schemes, much of which is still to be added to CUDA-Q. The addition of such schemes in future releases would further increase our shot efficiency. (Inset) the effect of additional GPUs on the intra-trajectory shot efficiency, with the multi-shot PTS algorithm benefiting substantially. As Pre-Trajectory Sampling is embarrassingly parallel, inter-trajectory efficiency scaling (not shown) is by definition linear with GPU number.}
\label{fig:tn_benchmarking}
\end{figure}

While the tensor network implementation of Pre-Trajectory Sampling methods remain limited by the current features of CUDA-Q, considerable speedups were still accessible. As displayed in Fig.\ \ref{fig:tn_benchmarking} (left axis, green), for an 85-qubit circuit magic state distillation preparation circuit simulated on four GPUs, an over $16x$ efficiency increase in shot collection was obtained for batched samples of just $10^3$ shots, a regime wherein nearly the entirety of shots collected would remain unique. In addition to the embarrassingly parallel nature of our method as applied independently to multiple trajectories at a time (inter-trajectory simulation), we note that we obtain a nearly linear speedup with increasing numbers of GPUs (NVIDIA H100 GPUs with 80GB of vRAM each) for intra-trajectory simulation as well (inset). We leave as future work the opportunity to further optimize the tensor network implementation for PTSBE but include discussion for speedup opportunities here. In particular, the efficiency of the tensor network implementation would be considerably improved via the addition of a few key features in CUDA-Q. Configurable tensor network contraction path optimization and caching, along with methods to allow small variations in cached paths, will allow for the calculation and storage of relatively few, highly efficient contraction paths. This will not only remove the contraction path finding overhead entirely on the per-trajectory basis, it will also decrease the computational overhead of the contraction itself from that of an average-case path to that of an optimized path, the speedup of which can reach multiple orders of magnitude. Likewise, the current sampling algorithm for tensor networks requires nearly all of the tensor network contraction process to reoccur for each sample, caching only the minimally optimized contraction path found for each round of shots. New contraction path methods based on conditional and correlated tensor network sampling will provide considerable speedup by reusing information from partial tensor network contractions, batching large numbers of shots more efficiently through the use of cached intermediates.

Our techniques were used to produce massive corpuses of noisy quantum data that is suitable for downstream tasks such as training an ML-based QEC decoder. Specifically, we collected one trillion 35-qubit shots and one million 85-qubit shots. Using PTSBE methods, these extremely large datasets can be obtained with unprecedented speed, requiring just $4,445$ H100 GPU hours and $2,223$ H100 GPU hours on NVIDIA's Eos DGX Superpod \cite{eos} for $10^6$ statevector and $100$ tensor network shots per batch, respectively. All experiments were done using CUDA-Q v0.10, with statevector and tensor network data points consisting of $100$ and $200$ experiments, respectively.

\section{Conclusion}

In this manuscript, we introduced PTSBE, a two part method for strategic quantum state sampling for noisy, universal quantum systems that not only provides fine-grained control over the error sampling process and valuable metadata on the origin of the errors incurred, it also permits highly optimized batched execution of many quantum shots without the repreparation of quantum states. Our method has demonstrated a $10^6$x and $16$x speedup for statevector and tensor network backends on useful, large-scale quantum error correction codes, enabling us to collect massive data corpuses of one trillion and one million quantum shots, respectively. While training ML-based QEC decoders is a primary target for our technique, we have established its potential for quantum applications in AI more generally. In subsequent works, further development of the tensor network backend will greatly increase the efficiency with which it can collect quantum data. This will enable massive data generation for very large, noisy, universal quantum computer simulations, a crucial frontier for the advancement of AI for QEC.



\bibliographystyle{ACM-Reference-Format}
\bibliography{main}

\end{document}